\documentclass[aps,twocolumn,showpacs]{revtex4}
\usepackage{psfig}

\def\order#1{{\cal O}\left(#1\right)}
\def\eq#1{(\ref{#1})}
\def\ep{\epsilon}
\newcommand{\be}{\begin{equation}}
\newcommand{\ee}{\end{equation}}
\newcommand{\ba}{\begin{eqnarray}}
\newcommand{\ea}{\end{eqnarray}}

\newcommand{\vecr}{\mbox{$\vec{r}$}}

\begin{document}

\title{
%
%
\[ \vspace{-2cm} \]
\noindent\hfill\hbox{\rm  } \vskip 1pt
\noindent\hfill\hbox{\rm Alberta Thy 07-01} \vskip 1pt
\noindent\hfill\hbox{\rm SLAC-PUB-8986} \vskip 1pt
\noindent\hfill\hbox{\rm hep-ph/0109054} \vskip 10pt
%
%
Charmonium decays: $J/\psi\to e^+e^-$ and $\eta_c \to \gamma\gamma$ 
}

\author{Andrzej Czarnecki}
\affiliation{
Department of Physics, University of Alberta\\
Edmonton, AB\ \  T6G 2J1, Canada\\
E-mail:  czar@phys.ualberta.ca}

\author{Kirill Melnikov}
\affiliation{Stanford Linear Accelerator Center\\
Stanford University, Stanford, CA 94309\\
E-mail: melnikov@slac.stanford.edu}

\begin{abstract}

We compute the ${\cal O}(\alpha_s^2)$ correction to the decay rate
$\eta_c \to \gamma \gamma$ and discuss its implications for precision
quarkonium physics. We study the suitability of the ratio
$\Gamma(J/\psi \to e^+e^-)/\Gamma(\eta_c \to \gamma \gamma)$, in which
the non-perturbative or soft effects cancel at
$\order{\alpha_s^{0,1}}$, for extracting fundamental parameters of QCD
at low energies.  We show that the QCD-based theory of charmonia is
not capable of predicting this ratio with any degree of confidence.

\end{abstract}

\maketitle

The physics of $c \bar c$ mesons (charmonium) is a mature field with a
long history. Discovery of the $J/\psi$ resonance at SLAC and
Brookhaven in the autumn of 1974 is often called ``November
revolution'', to emphasize its importance for the development of QCD
and of the Standard Model.  Already the early theoretical papers on
the subject interpreted the observed narrow resonance as the
non-relativistic bound state of $c \bar c$ quarks and thus
initiated studies of heavy quarks in QCD \cite{Appelquist:1975zd}.

During almost thirty years since the discovery of the first
charmonium, both experimental and theoretical studies of these mesons
have been successfully pushed forward.  The spectrum, lifetimes and
branching ratios have been precisely measured. Further progress is
expected at the planned dedicated facility CLEO-c
\cite{Briere:2001rn}. On the theoretical side, various attempts have
been made to improve the description of these hadrons.  In particular
a lot of effort went into determining how well the $c \bar c$ bound
states can be described if one starts directly from the QCD
Lagrangian.

An important recent development has been the introduction of effective
field theory techniques for describing hadrons consisting of two
non-relativistic heavy quarks \cite{Bodwin:1995jh,Caswell:1986ui}.
This effective field theory, the Non-relativistic Quantum
Chromodynamics (NRQCD), connects the original QCD and the new NRQCD
Lagrangian which takes full advantage of the fact that the quarks in
the $c \bar c$ bound state are non-relativistic.  As usual in
effective field theories, the two Lagrangians are matched
perturbatively at energy scales around the charm quark mass $m_c
\equiv m \sim 1.7~{\rm GeV}$.

In recent years the effective field theory approach to
non-relativistic bound states has been extended further.  It has been
noticed that two additional scales, the heavy quark momentum $mv$ and
the heavy quark binding energy $m v^2$, exist in quarkonia and, for
sufficiently heavy quarks, permit perturbative treatment.
Unfortunately, this is not quite possible for charmonium because the
$c$ quark mass is too small and therefore $mv \sim mv^2 \sim
\Lambda_{\rm QCD}$.  However, we still have $\Lambda_{\rm QCD} \ll m$
and therefore there is a chance that integrating out hard ($k \sim m$)
modes and matching QCD at NRQCD perturbatively is a sensible thing to
do.  If the soft effects are universal, they cancel in ratios of
various observables and a clean perturbative QCD prediction emerges.

For various observables, this approach has been taken
at order ${\cal O}(\alpha_s)$ and the common perception is 
that it works rather well. Let us consider the simplest decays 
of the ground state charmonia, $J/\psi \to e^+ e^-$ 
and $\eta_c \to \gamma \gamma$. To order ${\cal O}(\alpha_s)$
the decay rates can be written as
\ba
&&\Gamma_{\psi} \sim \psi^2(r=0) \left ( 1 
+ x_\psi \cdot \frac {\alpha_s}{\pi}   \right ), 
\nonumber \\
&&\Gamma_{\eta} \sim \psi^2(r=0) \left ( 1 
+ x_\eta \cdot \frac {\alpha_s}{\pi}  \right ), 
\ea
where $\psi(r)$ is the charmonium wave function and
$x_{\psi,\eta}$ are numbers which
can be determined by perturbative matching of the 
QCD and NRQCD Lagrangians.  Taking the ratio of these two decay 
widths, one obtains a prediction that is  free from any non-perturbative 
uncertainties,
\be
\frac {\Gamma_{\psi}}{\Gamma_{\eta}} \sim 1 
+ \left ( x_\psi - x_\eta \right ) \frac {\alpha_s}{\pi} ,
\label{eq:ratio}
\ee
and can be either compared to the data provided $\alpha_s(m)$ is known
or used to extract the value of $\alpha_s$.  For
example, the CLEO collaboration has  recently 
\cite{Brandenburg:2000ry} 
determined the strong coupling constant $\alpha_s(m_c)$
from the branching ratio of $\eta_c \to \gamma \gamma$  assuming that
the $\eta_c$ total decay width is saturated by the two-gluon channel.
Since the ratio \eq{eq:ratio} is independent of the wave  functions,
its comparison with measurements has often
been considered a solid test of perturbative QCD. 

In the past few years the development of techniques for
non-relativistic effective field theories has undergone an important
transition and we can now study the next order in the strong coupling
constant expansion.  Interestingly, in doing so one encounters new
conceptual difficulties.  The most important problem is that the hard
Wilson coefficient of the operator responsible for the decay in the
leading order becomes infrared divergent at two loops, which implies
that the wave function at the origin becomes scale dependent.  If this
scale dependence were the same for spin triplet $J/\psi$ and
singlet $\eta_c$ states, this would not pose a difficulty since it
would cancel in the ratio.  However, this is not the case and the
divergent parts of the Wilson coefficients are spin-dependent. This
immediately implies that with ${\cal O}(\alpha_s^2)$ accuracy the wave
functions at the origin of $J/\psi$ and $\eta_c$ become {\it
different} and therefore the ratio of the corresponding decay widths
is sensitive to some soft-scale effects. Since it is rather difficult
to compute these effects accurately, the QCD-based
prediction for the ratio of decay widths $\Gamma(J/\psi \to
e^+e^-)/\Gamma(\eta_c \to \gamma \gamma)$ becomes much less precise if
one computes higher order corrections in ${\cal O}(\alpha_s)$ --- a
somewhat paradoxical situation.  This is the principal message we
would like to get across in this Letter.

The remaining part of this Letter is organized as follows. We first
consider hard renormalization factors of the non-relativistic
operators responsible for the decays $\eta_c \to \gamma \gamma$ and
$J/\psi \to e^+e^-$.  We then show how the soft effects are taken into
account in our calculation and derive our final result for the ratio
of the decay rates of $J/\psi \to e^+ e^-$ and $\eta_c \to \gamma
\gamma$.

The hard renormalization factor for the
spin-singlet decay operator has been evaluated very recently
\cite{Czarnecki:2001gi}, extending an earlier QED result obtained
for the para-positronium decay
\cite{Czarnecki:1999gv,Czarnecki:1999ci}. 

\begin{figure}[htb]
\hspace*{-35mm}
\begin{minipage}{16.cm}
\begin{tabular}{cc}
\psfig{figure=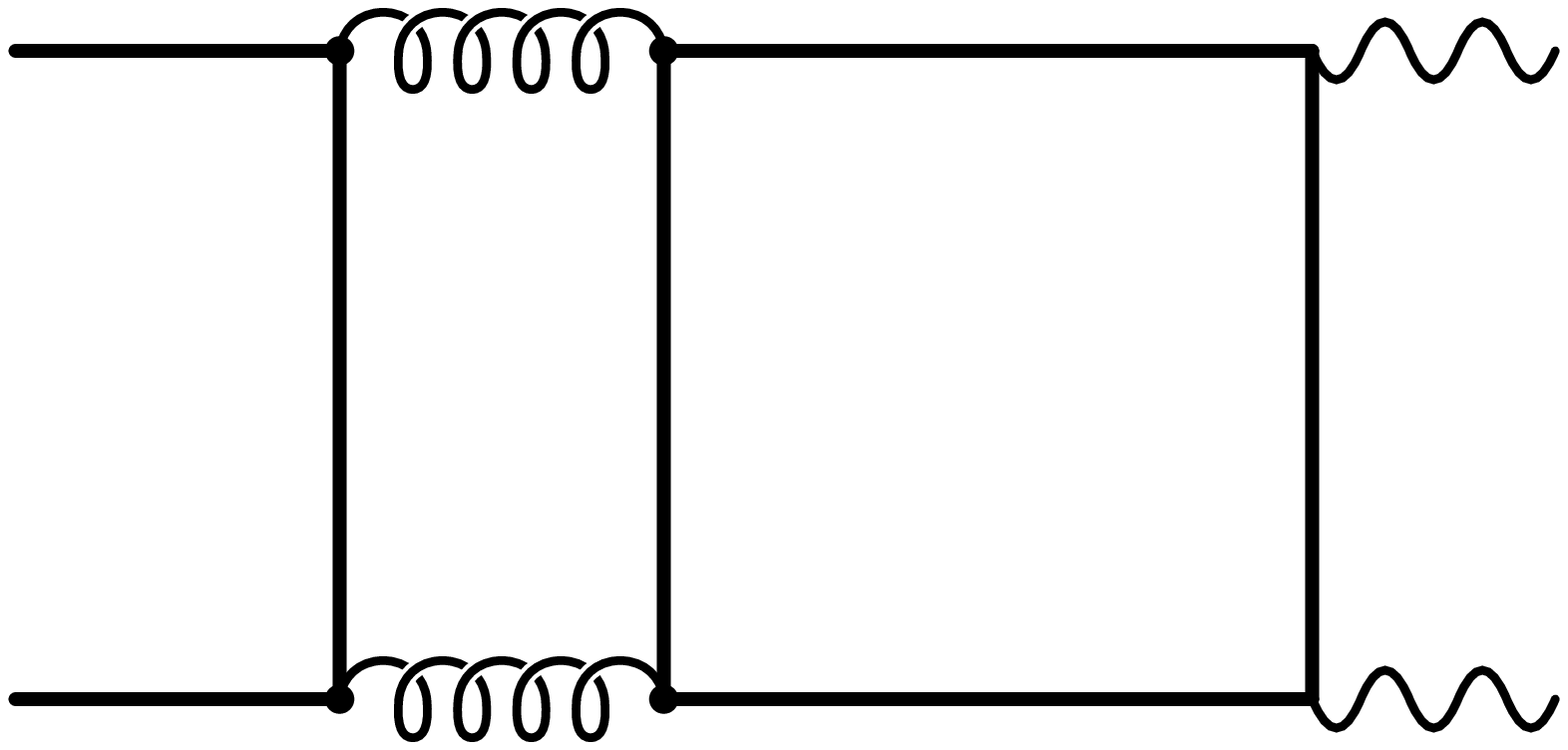,width=35mm}
&\hspace*{5mm}
\psfig{figure=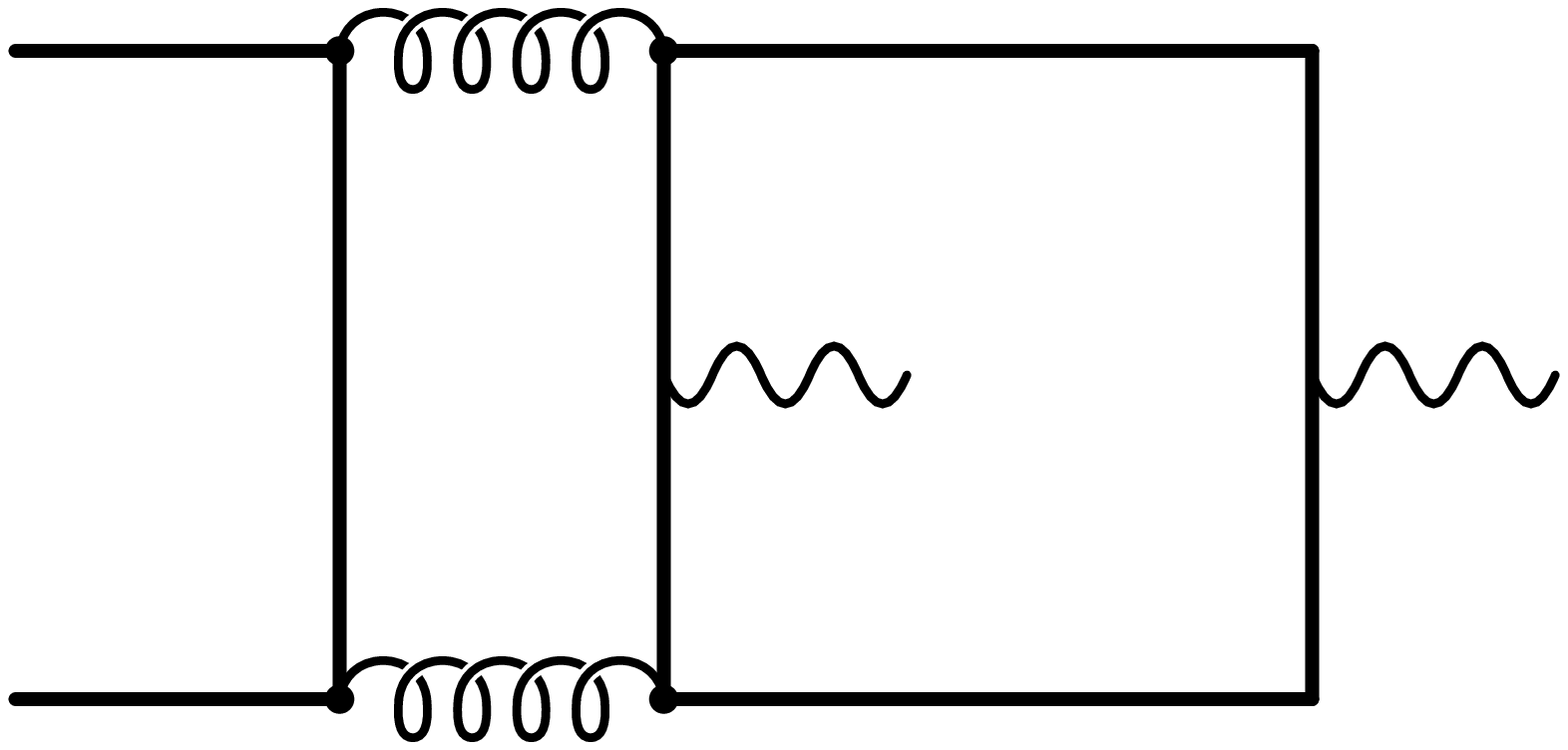,width=35mm}
\\[1mm]
\end{tabular}
\end{minipage}
\caption{Diagrams not considered in this paper: ``light-by-light''
scattering contributions.} 
\label{fig:light}
\end{figure}

We did not include the diagrams shown in Fig.~\ref{fig:light} in our
final result.  We have checked with a rough approximation that this
finite and gauge-invariant subset contributes only insignificantly.

Our result for the hard renormalization factor for the singlet decay
operator is (we use dimensional regularization with $D=4-2\ep$)
\ba
1-\left( {5\over 2} -{\pi^2\over 8}\right) C_F{\alpha_s\over \pi}
+ s_2(\mu) C_F\left({\alpha_s\over \pi}\right)^2,
\label{hard}
\ea
where $\alpha_s = \alpha_s(m)$ and 
\ba
s_2(\mu) &=& C_F s_A + C_A s_{NA} + N_L T_R s_L + N_H T_R s_H,
\nonumber \\
s_A(\mu) &=&  -21.0 -\pi^2\left(\frac{1}{4\ep} + \ln{\mu\over m} \right),
\nonumber \\
s_{NA}(\mu) &=&-4.79 - {\pi^2\over 2} \left(\frac{1}{4\ep} + \ln{\mu\over m} \right),
\nonumber \\
s_L &=&{41\over 36}-{13\over 144}\pi^2-{2\over 3}\ln 2-{7\over
24}\zeta_3
\simeq -0.565, 
\nonumber \\
  s_H &=& 0.22.
\ea
In the above equation $\mu$ is the factorization scale that separates 
relativistic and non-relativistic momenta in the NRQCD framework.

The matching coefficient for the vector current, relevant for the 
decay $J/\psi \to e^+e^-$,  can be found in Eqs.~(10-15) of 
Ref.~\cite{threshold} (see also \cite{matchBSS}). 
Divergences in the matching coefficients of the two currents are
different.  In the 
$\overline {\rm MS}$ renormalization scheme the
ratio of the decay widths is (we use $C_F={4\over 3},\, C_A=3,\,
T_R={1\over 2},\, N_L =3,\, N_H=1$) 
\ba
{\cal R} &\equiv &
 \frac{\Gamma(J/\psi\to e^+e^-)}{\Gamma (\eta_c \to \gamma\gamma)}
\nonumber \\
& = & \frac{1}{3 Q_c^2}  \left [ 
1-0.62~\alpha_s(m)
+\alpha_s^2 
\left ( 2.64 - 2.37 \ln \frac {m}{\mu}  \right ) \right ] 
\nonumber \\
&&\qquad \qquad
\times \left ( \frac {\psi^2_\psi(0)}{\psi^2_\eta(0)} \right )_\mu.
\label{hr}
\ea
Obviously, the wave functions at the origin become renormalization
scheme and factorization scale dependent and eq.~(\ref{hr}) indicates
that they must be different for $J/\psi$ and $\eta_c$.  
Since they can no longer be eliminated in the ratio of
the decay widths, one looses predictive power. Note also that
eq.~(\ref{hr}) involves the $\overline {\rm MS}$ renormalized wave
functions, which cannot be directly determined even on the lattice.
Let us stress that to reach these conclusions one only has to
integrate out relativistic degrees of freedom.  No assumption about
the  dynamics of the bound state (besides its quantum numbers) is
necessary.

To cancel the divergences of the Wilson coefficients we have to
calculate certain soft effects.  For charmonium, which is not a
Coulombic bound state, a completely ``honest'' calculation is not
possible.  However, we will show that with plausible assumptions one
can determine the soft contribution to the ratio of two decay widths
using experimental data on $e^+e^- \to c \bar c$.  Two types of
relativistic corrections have to be considered: to the amplitude and
to the wave function at the origin.  It turns out that the former are
the same for $J/\psi \to e^+e^-$ and $\eta_c \to \gamma \gamma$ and
cancel in the ratio (for this reason they were not included in
eq.~(\ref{hr})).  As for the latter, only spin-dependent effects can
survive in the ratio.  We assume that the only operator in the
non-relativistic Hamiltonian responsible for the hyperfine splitting
is 
\be
\delta H = - \frac {\pi \alpha_s C_F
}{4 d m^2} [ \sigma_i, \sigma_j] [\Sigma_i,\Sigma_j] \delta(\vecr),
\label{hfso}
\ee
where the Pauli matrices $\sigma$ and $\Sigma$ act respectively on
spins of the charm quark and of the antiquark.  This operator follows
from a one-gluon exchange diagram and its QED analog is the hyperfine
splitting operator in the Breit Hamiltonian. With this operator we
compute the wave functions at the origin,
\ba
&& \psi_\psi^2(0) = 1 +\ldots +\frac {2 C_F \alpha_s \pi}{m^2} 
\left (\frac {2}{3} + \frac {10}{9} \ep \right ) \widetilde G(0,0),
\nonumber \\
&& \psi_\eta^2(0) = 1+\ldots - \frac {2 C_F \alpha_s \pi}{m^2} 
\left (2 + 6 \ep \right ) \widetilde G(0,0),
\label{wfs}
\ea
where $\widetilde G(0,0)$ is the reduced Green's function of the
non-relativistic  $c \bar c$ state, computed at the ground state
energy $E_1$ (at this level we neglect the difference between $J/\psi$
and $\eta_c$ masses),
\be
\widetilde G(0,0) = {\sum\limits_{n}}' \frac {|\psi_n(0)|^2}{E_1 - E_n},
\label{gf}
\ee 
and the prime means that the sum does not include the ground state
$J/\psi$ ($n=1$).  The dots in eq.~(\ref{wfs}) indicate that
spin-independent corrections are also present but they drop out in the
ratio of the decay widths.

The Green's function at the origin is divergent and it is precisely the 
divergence needed to cancel that in the ratio of the two
hard Wilson coefficients. To illustrate this, we first 
consider an academic example of ultra-heavy quarkonia,  where  
binding effects can be computed  in the Coulomb approximation. 
In this case, the Green's function at the origin $\widetilde G(0,0)$ can be 
extracted from Ref.~\cite{Czarnecki:1999mw} and reads
\be
\widetilde G(0,0) =  -{C_F \alpha_s m^2 \over 4\pi }
\left({1\over 4\epsilon} + \frac{3}{2} 
-\ln \frac {{m\alpha_s C_F}}{\mu} \right).
\ee
We now substitute this result in to  eq.~(\ref{wfs}) and obtain 
the ratio of the decay widths of the spin-triplet and spin-singlet 
ultra-heavy quarkonia,
\ba
{\Gamma_{Q \bar Q \to e^+e^-} \over \Gamma_{Q \bar Q \to \gamma\gamma}}
&=& {1\over 3Q_c^2}
\left[ 
1 - 0.62 a_s + a_s^2 \left( 2.37 \ln a_s - 1.8 \right)
\right],
\nonumber \\
\label{resac}
\ea 
where $a_s = \alpha_s(m_Q)$. The coefficient of the $a_s^2 \ln a_s$ 
term, $4C_F^2/3\simeq 2.37$,  agrees with  \cite{Kniehl:1999mx}. 

We now turn to the more difficult case of charmonium, where one can
not use the Coulomb approximation for the low energy dynamics. We will
still use the hyperfine splitting operator, eq.~(\ref{hfso}), to
describe the hyperfine splitting; for this reason eqs.~(\ref{wfs}) are
still valid.  The challenge is to compute the Green's function
$\widetilde G(0,0)$ in dimensional regularization without invoking the
Coulomb approximation.

We will perform such computation in two different ways dealing either
with the observed cross section $e^+e^- \to c \bar c$ or with a simple
potential model constructed to describe charmonium.  It will be clear
that both calculations could be improved.  Here, our primary goals are
to demonstrate how they can be carried out using simple approximations
and to give reasonable estimates of the second order
corrections to the ratio of the two decay widths.

We first describe the calculation which utilizes experimental 
data on $e^+e^- \to c \bar c$. To this end, we  separate 
the Green's function into a contribution of charmonium  resonances 
and that of the continuum,
\ba
\widetilde G(0,0) = G^{\rm res}(0,0) + G^{\rm cont}_{E_1}(0,0).
\ea
 The former is finite and can be computed
using available data on the spectrum and $e^+e^-$ decay widths of the
spin 1 resonances, 
\ba
\widetilde G^{\rm res}(0,0)& =& \frac {1}{16Q_c^2 \pi \alpha^2}
{\sum_n}' \frac {M_n^2 \Gamma_{\psi(n) \to e^+ e^-} }{M_1 - M_n}
\nonumber \\
& \approx& -0.073 \mbox{ GeV}^2,
\label{reson}
\ea
where we have employed the mass and width information on the first six
$\psi$ resonances \cite{PDG2000}.

The continuum contribution is, on the other hand, divergent. To
determine this divergence using as little input information as
possible, we proceed in the following way.  We consider
continuum contribution as a function of $E$, 
\be 
\widetilde G^{\rm cont}_E(0,0) 
     = {\sum\limits_{E_n > 0}}' \frac {|\psi_n(0)|^2}{E - E_n}, 
\ee 
and take the derivative with respect to $E$. 
We then solve the resulting differential equation and obtain
\be 
\widetilde G^{\rm cont}_{E_1} (0,0) 
= \widetilde G^{\rm cont}_{E_i} + \int \limits_{E_i}^{E_1} 
{\rm d}E 
\frac{\rm d }{{\rm d} E} \widetilde G^{\rm cont}_E(0,0).  
\label{intrep}
\ee 
We can further use the relation between $R_c = 
\sigma(e^+e^- \to c \bar c)/\sigma(e^+e^- \to \mu^+ \mu^-)$ 
and the Green's functions $\widetilde G(0,0)$ to relate eq.~(\ref{intrep}) 
to experimental data. We obtain
\be 
\widetilde G^{\rm cont}_{E_1} (0,0) 
= \widetilde G^{\rm cont}_{E_i}
- \frac {m^2}{8\pi^2} \int \limits_{0}^{\infty} {\rm d}E~K(E)~R_c (E),
\label{intrep2}
\ee
where the function $K(E)$ is given by
\be
K(E) = \frac {E_1 - E_i}{(E-E_1)(E-E_i)}. 
\ee

The divergence in $\widetilde G(0,0)$ now resides in $\widetilde
G^{\rm cont}_{E_i}$ and the integral in eq.~(\ref{intrep2}) is
finite. For the initial condition $\widetilde G^{\rm cont}_{E_i}$ we
can choose a ``deep Euclidean'' point $E_i \to -\infty$, where
perturbative calculations are justified and where $\widetilde G^{\rm
cont}_{E_i}$ can be determined with, in principle, arbitrary
precision.  We therefore see that the divergent part of the Green's
function can be extracted from perturbative calculations and the
finite part can be obtained from $R_c$ (other options are to
compute the finite part using a potential  model for quarkonium or
NRQCD on the lattice).  This separation solves the problem in
principle and provides a way to determine the $\overline {\rm MS}$
charmonium wave functions without assuming that the bound state is 
Coulombic; the only true model-dependence remaining in this
calculation is the form of the operator responsible for the hyperfine
splitting, eq.~(\ref{hfso}).

The data on $R_c$ is not  quite precise yet.  
Its high energy asymptotics (in the 
non-relativistic sense) is fixed since the dependence on the initial 
energy $E_i$ in eq.~(\ref{intrep2}) should cancel. 
We therefore  write
\be
R_c(E) = 2 \left ( \sqrt{ \frac{E}{m} } + \frac{\pi C_F \alpha_s}{2} \right )
+R_c^{\rm npt}(E).
\label{anrc}
\ee
For $R_c^{\rm npt}(E)$ we choose:
\be
R_c^{\rm npt}(E) = -\pi C_F \alpha_s \theta(E_0 - E),
\ee
with $E_0 = \sqrt{s_0} - 2m$ and $\sqrt{s}_0 = 4~{\rm GeV}$. This ansatz 
is motivated by the data on $R_c$ in \cite{Zhao:2000bv} where one sees 
that there is no need for a large second term in eq.~(\ref{anrc}) 
below $4$ GeV. On the other hand we do need this term at higher 
energies, since otherwise ``perturbative'' and ``non-perturbative''
expressions do not match.

We obtain
\ba
\lefteqn{\widetilde G^{\rm cont}_{E_1}(0,0) =  \frac{m^2}{4\pi}\left[
\sqrt{\frac{-E_1}{m}} 
 - C_F \alpha_s  \cdot
\right.  }
\nonumber \\
&& 
\left.
\cdot
\left( {1\over 4\epsilon}
- \ln \sqrt{ \frac {-4mE_1}{\mu ^2} } + \frac{1}{2} 
- \frac{1}{2} \ln \frac{E_0 - E_1}{E_1}
\right)
\right].
\label{contin}
\ea

The soft contributions to the decay width ratio are obtained by
employing eqs.~(\ref{reson}) and (\ref{contin}) in eq.~(\ref{wfs}).
Before combining them with the hard
contributions in \eq{hr}, we note that the BLM effects \cite{BLM} were
computed in Ref.~\cite{braaten} for the rates of $J/\psi \to
e^+e^-$ and $\eta_c \to \gamma \gamma$.  These corrections turn out to
have different signs and are enhanced in the ratio. For this reason,
we decided to eliminate them by choosing different scales for the
strong coupling constant in the Wilson coefficients for $J/\psi \to
e^+e^-$ and $\eta_c \to \gamma \gamma$. We then find
\ba
\lefteqn{
{\Gamma_{J/\psi\to e^+e^-} \over \Gamma_{\eta_c \to \gamma\gamma}}
= {1\over 3Q_c^2}\left[ 1 - 1.7~a_{S=1} + 1.1~ a_{S=0} 
 \phantom{1\over 1} 
\right.
}
  \nonumber \\
&&\hspace*{-5mm} \left. 
 + \alpha_s^2
\left(1.19 \ln \frac{E_1-E_0}{m} 
 + 3.66 + 1.8 \sqrt{\frac{-E_1}{m \alpha_s^2}}
 - \frac {1.64}{\alpha_s m^2}
\right )
\right],
\nonumber \\
\label{res}
\ea
where $a_{S=0} = \alpha_s(1.95~m)$ and $a_{S=1} = \alpha_s(0.63~m)$
and the charm mass should be expressed in GeV.
The scale of the coupling constant in the second order correction 
is not specified;  we will use $\alpha_s = 0.3$ for the estimates. 
The inverse power of $\alpha_s$ in the square brackets arises because 
we have used experimental data to compute the contribution 
of the resonances to 
$\widetilde G(0,0)$ and also used the energy of the ground state 
$E_1 = M_{J/\psi} -2m$ to estimate the continuum contribution.
This spoils the homogeneity  in $\alpha_s$.

\begin{figure}[htb]
\vspace*{20mm}
\hspace*{-60mm}
\begin{minipage}{16.cm}
\begin{picture}(100,100)
\put (190,-0) {$m$ [GeV]}
\put (-10,125) {$f_2(m)$}
\hspace*{5mm}
\psfig{figure=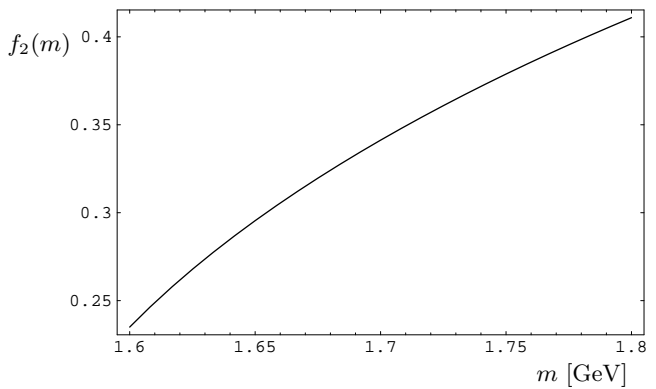,width=80mm}
\end{picture}
\end{minipage}
\caption{The second order correction in eq.~(\ref{res1}) as a function of
the charm quark mass.}
\label{fig:mc}
\end{figure}

Employing, for the sake of illustration, 
$\alpha_s=0.3$, $\alpha_s(0.63~m) = 0.35$ and 
$\alpha_s(1.95~m) =0.26$ in eq.~(\ref{res}), one finds 
\ba
{\Gamma_{J/\psi\to e^+e^-} \over \Gamma_{\eta_c \to \gamma\gamma}}
&=& {1\over 3Q_c^2}\left[ 1 - 0.32 + f_2(m) + \order{\alpha_s^3}
\right],
\nonumber\\
\label{res1}
\ea 
where the three terms in the brackets are the tree level, the ${\cal
O}(\alpha_s)$ and the ${\cal O}(\alpha_s^2)$ corrections,
respectively. As shown in Fig.~\ref{fig:mc}, the second order
correction depends strongly on the value of the charm quark mass.
It is large and very sensitive to the form of 
$R_c^{\rm npt}$ assumed in eq.~(\ref{anrc}). 
For example, if we use the Coulomb approximation to estimate the 
value of the wave function in the continuum, we obtain a
negative result for the second order correction.
Within our model for the continuum, we estimate
\ba
{\Gamma_{J/\psi\to e^+e^-} \over \Gamma_{\eta_c \to \gamma\gamma}}
&=& {1\over 3Q_c^2}\left[ 1 - 0.32 + 0.3 \pm 0.1 + \order{\alpha_s^3}
\right].
\nonumber \\
\label{res2}
\ea 

As we mentioned earlier, another possibility to obtain the Green's 
function necessary to compute the ratio of two decay widths is 
to use either QCD on the lattice or potential models. Here we 
would like to illustrate this point by considering a simple 
potential model. This will allow us to check that the estimate, 
eq.~(\ref{res2}), is reasonable. Although the potential 
model below is really simple, the calculation can be repeated with 
more sophisticated potentials, provided that, at short distances, 
these potentials match the QCD analog of the Coulomb potential. 

To describe charmonium, we will use the Schr\"odinger equation 
with the potential
\be
V(r) = -\frac{C_F \alpha_s}{r} +br +V_0,
\ee
where $b = 0.18~{\rm GeV}^2$ and $V_0$ is adjusted  to give
the correct mass of the $1S$ state ($J/\psi$) for given values 
of $m$ and $\alpha_s$. 

To compute the Green's function $G(0,0)$ we solve the 
Schr\"odinger equation following the treatment in 
\cite{Strassler:1991nw} and obtain the following representation 
for the full (including the ground state) Green's function:
\be
G_E(0,0) = \lim _{r \to 0} -\frac{m}{4\pi} \left [ \frac{1}{r} - 
C_F \alpha_s m \ln(r) + B(E) \right ],
\ee
where the function $B(E)$ is derived from the large $r$ limit of the 
ratio
of two solutions of the Schr\"odinger equation with prescribed 
behavior at the origin (see \cite{Strassler:1991nw}). In order to obtain 
the Green's function in dimensional regularization, which is needed 
for our purposes, we write
\ba
G_E(0,0) &\equiv & G_E - G_{E_1} + G_{E_1} 
\nonumber \\
&=&-\frac{m}{4\pi} B(E) + \frac{m}{4\pi} B(E_1) + G_{E_1}. 
\ea
We then take the limit $E_1 \to -\infty$,  perturbatively 
compute the Green's function $G_{E_1}$
and derive
\ba
\lim_{E_1 \to -\infty} \frac{m}{4\pi} B(E_1) + G_{E_1} =
&&\hspace*{-4mm}
-\frac{m^2 C_F \alpha_s}{4\pi} \left (\frac {1}{4\ep} - \ln
\frac{m}{\mu}   
\right ) + c,
\nonumber \\
&&
\ea
where $c$ is
\ba
c &=& \lim _{E_1 \to -\infty} \left [  \frac{m}{4 \pi} B(E_1) + 
\frac{m^2}{4\pi} \sqrt{\frac{-E_1}{m}} 
\right. 
\nonumber \\
&& \left.
- \frac{m^2 C_F \alpha_s}{4\pi} 
\left ( - \ln \sqrt{\frac{-4 E_1}{m}} + \frac{1}{2} \right ) \right ]
\nonumber \\
&=& -\frac {m^2 C_F \alpha_s}{8 \pi} \left (-1+2 \gamma_E +2 \ln m \right ), 
\ea
and $\gamma_E$ is the Euler constant.

In order to compute the reduced Green's function $\widetilde G(0,0)$
eq.~(\ref{gf}), we find the first eigenvalue numerically and remove
the pole from $B(E)$. 

We have computed the ratio of two decay widths for different values of 
$\alpha_s$ and the mass of the charm quark. We obtain
\ba
{\Gamma_{J/\psi\to e^+e^-} \over \Gamma_{\eta_c \to \gamma\gamma}}
&=& {1\over 3Q_c^2}\left[ 1 - 0.32 + 0.20 \pm 0.05 
+ \order{\alpha_s^3} \right].
\nonumber \\
\label{resm}
\ea
We see that the result of the potential model calculation 
is relatively close to the result of eq.~(\ref{res2}).  The advantage
of the potential model calculation is its fair stability 
against variations in $\alpha_s$ and 
$m$. We take the result in eq.~(\ref{resm}) as our final estimate.

In spite of the fact that the model leading to eq.~(\ref{resm}) is
quite simple, we believe that eq.~(\ref{resm}) is important in
that it clearly shows the magnitude of second order QCD corrections
one might expect for such observables. 

It is interesting to note that there is a strong cancellation between
the first and second order effects in eqs.~(\ref{res2},\ref{resm}).
Neglecting all the radiative corrections and using $\Gamma_{\psi \to
e^+e^-} = 5.26~{\rm keV}$, we derive $\Gamma_{\eta_c \to \gamma
\gamma} = 7.01~{\rm keV}$, rather close to the central value reported
by CLEO collaboration \cite{Brandenburg:2000ry} $\Gamma_{\eta_c \to
\gamma \gamma}^{\rm exp} = [7.06 \pm 0.8({\rm stat}) \pm 0.4({\rm
sys}) \pm 2.3 ({\rm br})]~{\rm keV}$.

We conclude that eqs.~(\ref{res},\ref{res1},\ref{resm}), 
the principal results of this Letter,
illustrate an unexpected problem in the theory of heavy quarkonia at
the two-loop level.  In recent years we have learned how to integrate
out relativistic degrees of freedom efficiently and it seemed as if we
could improve  the accuracy of our predictions. This turns out not
to be the case.  The reason is that at ${\cal O}(\alpha_s^2)$ the soft
and relativistic effects do not decouple completely, as it happens at
${\cal O}(\alpha_s)$, and therefore, in general,  one cannot avoid
non-perturbative effects by taking ratios of different observables.
We have shown how, in principle, the soft contribution can be estimated
using experimental data or potential models. 

With the QCD corrections  as big as in eq.~(\ref{resm}), the
determination of $\alpha_s(m_c)$ from charmonia decay rates, as
e.g. in Ref.~\cite{Brandenburg:2000ry}, does not look trustworthy,
regardless of the fact that the numerical values of $\alpha_s$ turn
out to be in a theoretically sensible range. On the other hand,
it is interesting to point out that the ratio of the decay rates 
of $\eta_c \to \gamma \gamma$ to $\eta_c \to g g$, actually used 
in Ref.~\cite{Brandenburg:2000ry} for determination of $\alpha_s$,
is free from the soft effects we  discussed in this Letter, 
since it refers to the same initial state. It would therefore be 
interesting to compute second order QCD corrections to this ratio 
since in this case the  hard corrections alone might provide an
unambiguous answer.

Among various charmonium decays, only $J/\psi \to e^+ e^-$ and $\eta_c
\to \gamma \gamma$ have now been studied to ${\cal O}(\alpha_s^2)$.
Clearly these are the two simplest channels since they do not involve
any complications related to the dynamics of hadrons in the final
state.  If the understanding of even those simplest decays encounters
such difficulties, one should exercise great care when extracting
physical information from more complicated charmonium decays. The fact
that perturbative QCD appears to work well in the one-loop order is
certainly insufficient to ensure that heavy quarkonia are well
understood.

{\em Acknowledgments:} We are grateful to M.~B.~Voloshin for helpful
advice and to A. S. Yelkhovsky for helpful
discussions and sharing with us his
numerical algorithm for determining Green's functions.  
This research was supported in part by the Natural
Sciences and Engineering Research Council of Canada and by the DOE
under grant number DE-AC03-76SF00515.


\begin{thebibliography}{10}

\bibitem{Appelquist:1975zd}
T. Appelquist and H.~D. Politzer, Phys. Rev. Lett. {\bf 34},  43  (1975).

\bibitem{Briere:2001rn}
R.~A. Briere {\it et~al.}, preprint CLNS-01-1742 (unpublished).

\bibitem{Bodwin:1995jh}
G.~T. Bodwin, E. Braaten, and G.~P. Lepage, Phys. Rev. {\bf D51},  1125
  (1995).

\bibitem{Caswell:1986ui}
W.~E. Caswell and G.~P. Lepage, Phys. Lett. {\bf B167},  437  (1986).

\bibitem{Brandenburg:2000ry}
G. Brandenburg {\it et~al.}, Phys. Rev. Lett. {\bf 85},  3095  (2000).

\bibitem{Czarnecki:2001gi}
A. Czarnecki and K. Melnikov, hep-ph/0108233 (unpublished).

\bibitem{Czarnecki:1999gv}
A. Czarnecki, K. Melnikov, and A. Yelkhovsky, Phys. Rev. Lett. {\bf 83},  1135
  (1999), erratum: ibid. {\bf 85}, 2221 (2000).

\bibitem{Czarnecki:1999ci}
A. Czarnecki, K. Melnikov, and A. Yelkhovsky, Phys. Rev. A {\bf 61},  052502
  (2000), erratum: ibid. {\bf 62}, 059902 (2000).

\bibitem{threshold}
A. Czarnecki and K. Melnikov, Phys. Rev. Lett. {\bf 80},  2531  (1998).

\bibitem{matchBSS}
The matching coefficient for the vector current was also derived in M. Beneke,
  A. Signer and V. A. Smirnov, Phys. Rev. Lett. {\bf 80}, 2535 (1998).

\bibitem{Czarnecki:1999mw}
A. Czarnecki, K. Melnikov, and A. Yelkhovsky, Phys. Rev. {\bf A59},  4316
  (1999).

\bibitem{Kniehl:1999mx}
B.~A. Kniehl and A.~A. Penin, Nucl. Phys. {\bf B577},  197  (2000).

\bibitem{PDG2000}
D.~E. {Groom \em et al. (Particle Data Group)}, Eur. Phys. J. {\bf C15},  1
  (2000).

\bibitem{Zhao:2000bv}
Z.~G. Zhao, hep-ex/0012038 (unpublished).

\bibitem{BLM}
S.~J. Brodsky, G.~P. Lepage, and P.~B. Mackenzie, Phys. Rev. {\bf D28},  228
  (1983).

\bibitem{braaten}
E. Braaten and Y. Chen, Phys. Rev. {\bf D57},  4236  (1998), erratum: {\it
  ibid.} {\bf 59}, 079901 (1999).

\bibitem{Strassler:1991nw}
M.~J. Strassler and M.~E. Peskin, Phys. Rev. {\bf D43},  1500  (1991).

\end{thebibliography}

\end{document}